\documentclass[a4paper,11pt]{article}
\pdfoutput=1 

\usepackage{jinstpub} 
\def\fbi{\mbox{fb$^{-1}$}}%

\title{\boldmath Background in the CMS muon detectors: simulation 
and measurements with pp collision data}

\author[a]{Silvia Costantini}

\affiliation[a]{CERN, CH-1211 Geneva 23, Switzerland}

\emailAdd{silvia.costantini@cern.ch}

\abstract{
The CMS muon system at the LHC is built of different detector technologies.
The measurement of the background hit rates in the different muon detectors 
during the LHC Run-2 is of prime importance for an assessment of the longevity of the chambers 
and their on-board electronics, and therefore for the expected performance of the system at HL-LHC. 
Moreover, an accurate modelling of the backgrounds using simulations is also critical, 
as an estimation of the expected radiation background for the Phase-2 upgrade
can only rely on Montecarlo-based predictions in the regions where new detectors are being installed.
The present understanding of the backgrounds measured with data collected during the LHC Run-2, 
as well as at CERN high-intensity gamma irradiation facility, GIF++, is discussed,
together with the work made to improve the accuracy of the background modelling in Fluka and GEANT4 simulations.
}


\collaboration[c]{on behalf of CMS Collaboration}

\proceeding{iWoRiD 2019, 21$^{\text{th}}$  International Workshop on Radiation 
Imaging Detectors,
7-12 Jul 2019, Kolymbary, Chania, Crete (Greece)}

\begin{document}
\maketitle
\flushbottom



\section{The CMS muon system and its upgrade}
\label{sec:intro}

\begin{figure}[htbp]
\centering 
\includegraphics[width=.6\textwidth]{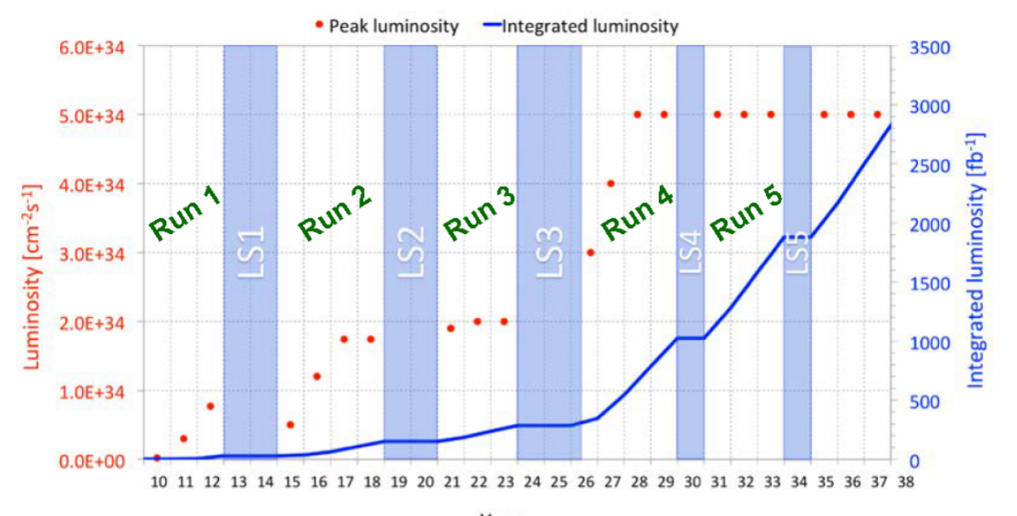}
\caption{\label{fig:fig1_lumi} 
Development of the instantaneous and integrated 
luminosities vs. time for the design HL-LHC. The present ``Phase-1'' data taking period will end in 2023, 
followed by a shutdown for the HL-LHC upgrade. The high luminosity data taking period, ``Phase-2'', 
is expected to last from 2026 till 2038.
}
\end{figure}

The High Luminosity LHC (HL-LHC) upgrade~\cite{tdr} is expected to increase the sensitivity to
new physics by increasing the center-of-mass energy for proton-proton (pp) collisions 
to 14 TeV and the instantaneous luminosity to $5 \cdot 10^{34}$ cm$^{-2}$ s$^{-1}$
($7.5 \cdot 10^{34}$ cm$^{-2}$ s$^{-1}$ in the ultimate scenario), with an average number 
of additional pp interactions (pileup) per bunch-crossing
expected to reach 140 (200 in the ultimate scenario). 
The integrated luminosity will increase in the coming two decades
from the design value of 300~\fbi of the present LHC ``Phase-1'', 
to 3000~\fbi (or to 4000~\fbi, expected ultimate value) 
during the so-called LHC ``Phase-2'', as shown in figure~\ref{fig:fig1_lumi}.

\begin{figure}[htbp]
\centering 
\includegraphics[width=.6\textwidth]{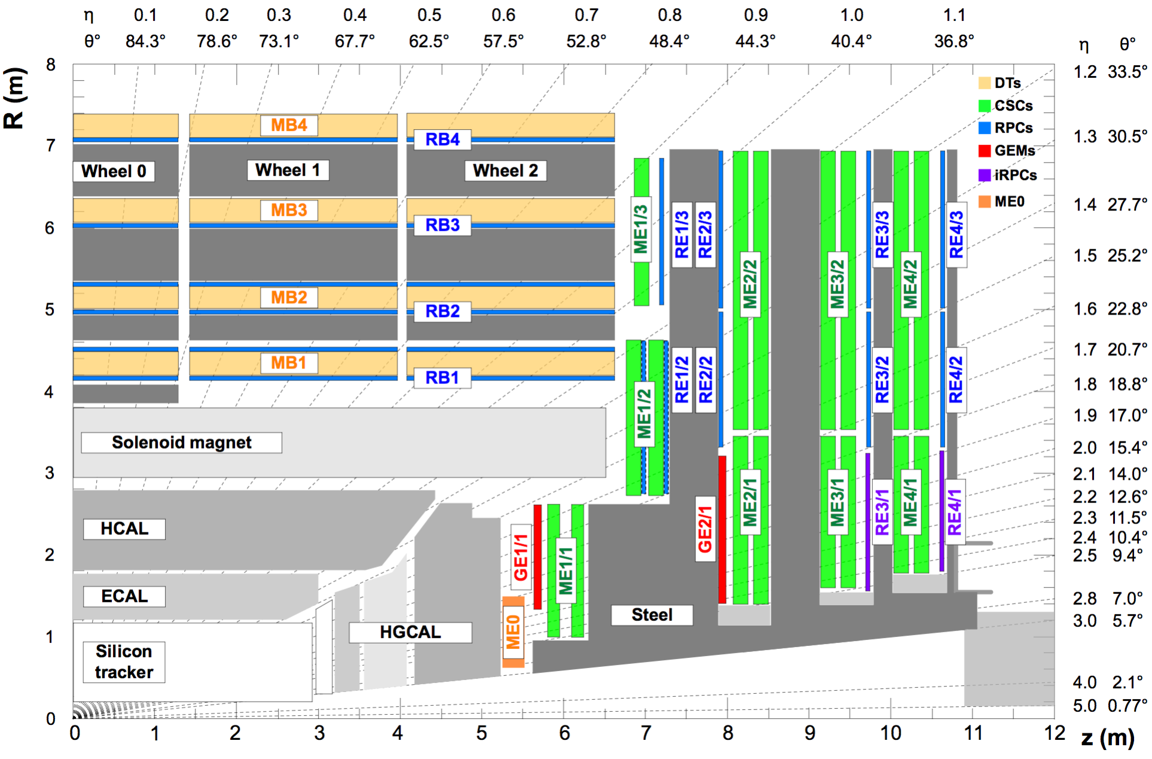}
\caption{\label{fig:quadrant} 
Schematic view, in the R-z plane, of one quadrant of the CMS detector,
with the axis parallel to the beam (z) running horizontally and the radius (R) 
increasing upward. The interaction region is at the lower left corner.
The position of the present RPC chambers is shown in blue. The RPCs
are both in the barrel (``RB'' chambers) and in the endcaps (``RE'') of CMS. 
The DT chambers are 
labeled ``MB''(``muon barrel'') and the CSC chambers are 
labeled ``ME'' (``muon endcap'').
The steel disks are displayed as dark gray areas.
Also shown in orange and red are the chambers of the proposed upgrade scenario
of the CMS Muon system, including Gas Electron Multiplier detectors
(labeled ``ME0'' and ``GE'') and iRPCs (``RE3/1'' and ``RE4/1'').  
}
\end{figure}

The detailed description of the CMS detector, together with a definition
of the coordinate system used, can be found in Ref.~\cite{cms_det}.
Muons with pseudorapidity in the range $| \eta | < 2.4 $ are measured 
in gas-ionization detectors embedded in the steel return yoke,
with detection planes made of three technologies:
Drift Tube chambers (DTs) and Cathode Strip Chambers (CSCs) 
serve as tracking and triggering detectors respectively in the barrel and the endcaps 
of the spectrometer, whereas Resistive Plate Chambers (RPCs) complement the DTs and CSCs 
both in the barrel and in the endcaps, and are mostly used in the trigger. 
In addition, multiple layers of Gas Electron Multiplier (GEM)~\cite{gem} chambers are being installed 
in the muon system endcaps at different stages of the CMS upgrade programme. 
The present CMS muon system, together with the detectors of the proposed HL-LHC
scenario, is shown in figure~\ref{fig:quadrant} and described in
detail in~\cite{muon}.


\section{Expected background in the CMS muon system}

The estimation of the dose and rate expected in the muon system from radiation background 
is essential to choose the appropriate detector technologies for the upgrade,
and to design the detectors and electronics that will be able to cope 
with the HL-LHC conditions.
The new detector technologies (the GEMs and the so-called ``improved RPCs'', referred to as
iRPCs in the following) that are to be installed
as part of the muon system upgrade, are designed to maintain excellent performance
throughout the whole HL-LHC operation.
Both the new and the
existing detectors (DTs, CSCs, RPCs)  are being tested (Section~\ref{sec:gif}) to demonstrate,
or confirm, their radiation hardness at the level required  to handle
the HL-LHC background conditions.

\begin{figure}[htbp]
\centering 
\includegraphics[width=.5\textwidth]{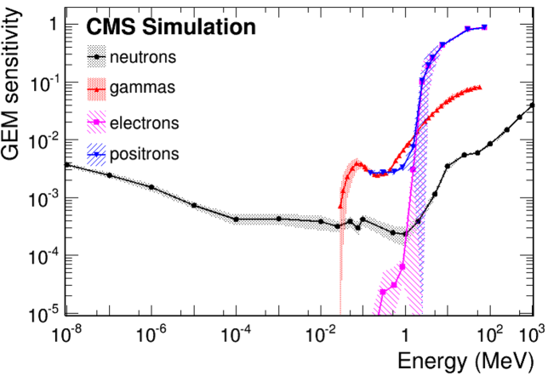}
\caption{\label{fig:sens} GEM sensitivities in the GE1/1 chambers as a function of the 
particle energies for neutrons, photons, electrons and positrons. The sensitivities
are simulated with GEANT4.
}
\end{figure}

\begin{figure}[htbp]
\centering 
\includegraphics[width=.5\textwidth]{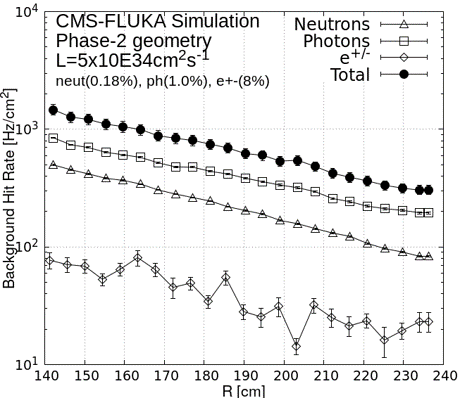}
\caption{\label{fig:exp_rates} 
Expected hit rates, at the HL-LHC design luminosity, in the GE1/1 chambers. Rates
values are determined normalizing fluxes estimated with FLUKA via average sensitivities
obtained with GEANT4 and shown for instance in figure~\ref{fig:sens}.}
\end{figure}

Proton-proton primary interactions and particle transport through the
CMS detector are simulated using FLUKA~\cite{fluka1,fluka2}, which is also used
to estimate the expected fluences, doses and fluxes from neutrons, photons and charged particles.
Hit rates are in turn estimated by normalizing the fluxes by the detector 
sensitivities determined with GEANT4~\cite{geant1,geant2,geant3}. Very accurate descriptions 
of the CMS detector and the CMS cavern are implemented in the FLUKA and
GEANT4 simulations, well reproducing the Run1 and Run2 measurements 
as shown in the next Section.
As an example, figures~\ref{fig:sens} and~\ref{fig:exp_rates}
show typical detector sensitivities
in the GEM chambers as a function of the particle energies, 
together with the resulting expected fluxes in the detector.

The DTs, CSCs and RPCs are replacing for Phase-2 their front-end electronics
in the chambers exposed to the largest background with new electronics
tested to cope with the expected neutron fluences and total ionization doses~\cite{tdr}.

\section{Background measurements}

\begin{figure}[htbp]
\centering 
\includegraphics[width=.5\textwidth]{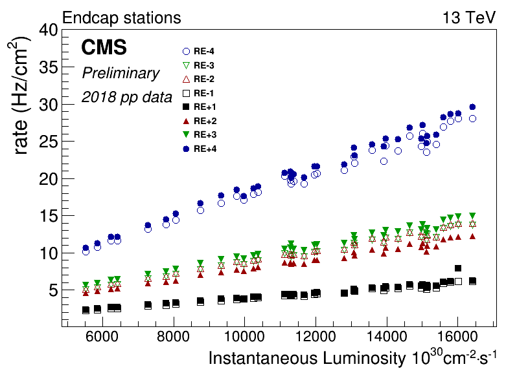}
\caption{\label{fig:fig3endcap} Rates measured in the endcaps of the RPCs as a function of the
instantaneous luminosity, at a center-of-mass energy of 13 TeV,
for values of the instantaneous luminosities up to 
$\approx 1.6 \cdot 10^{34}$~cm$^{-2}$~s$^{-1}$.
}
\end{figure}

\begin{figure}[htbp]
\centering 
\includegraphics[width=.8\textwidth]{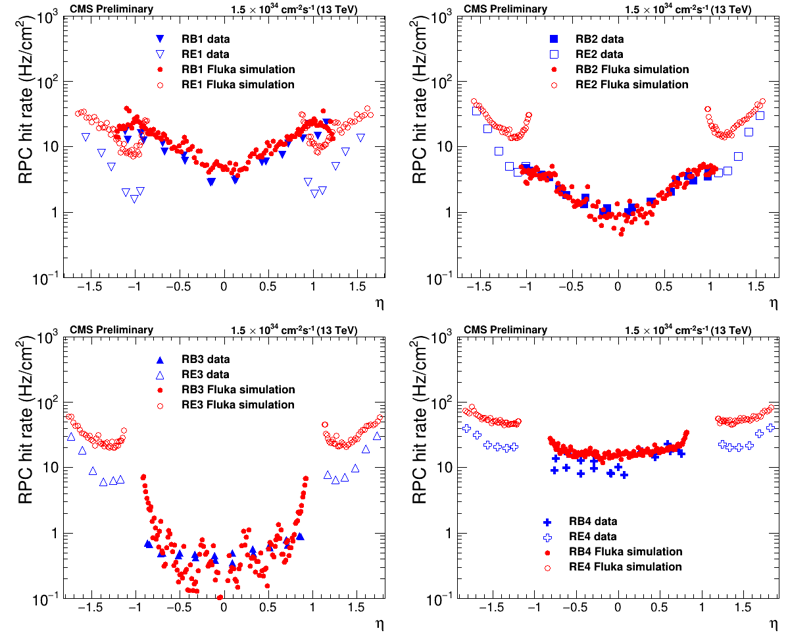}
\caption{\label{fig:fig3barrelendcap} 
Comparison between measurements and simulation in the barrel
and endcap stations of the RPC detector. The stations RB1-RE1 (top left),
RB2-RE2 (top right), RB3-RE3 (bottom left) and RB4-RE4 (bottom right)
are shown, respectively.
The simulation is able to reproduce
the shape of the data, with a quantitative agreement at the level of 10\%
to 30\% in the barrel, and within a factor 2 in the endcap. }
\end{figure}

The data collected during the LHC Run1 and Run2 have offered the opportunity to extensively
study the radiation background inside the CMS cavern and in the Muon system. A linear dependence
as a function of the LHC instantaneous luminosity has been observed by all subdetectors
over several orders of
magnitude of the instantaneous luminosity, as shown as an example in figure~\ref{fig:fig3endcap}.
The highest rates are observed in the external endcap chambers, more exposed to the neutrons
permeating the CMS cavern. 
Assuming that the linear dependence on the luminosity, observed from 
$\approx 10^{29}$~cm$^{-2}$~s$^{-1}$ to $\approx 10^{34}$~cm$^{-2}$~s$^{-1}$, will hold
at $5 \cdot 10^{34}$~cm$^{-2}$~s$^{-1}$, these measurements can be extrapolated
to estimate the expected background rates at the HL-LHC in the existing
detectors. 
Figure~\ref{fig:fig3barrelendcap} shows the RPC measurements in
the barrel and endcap compared to the expected hit rates in the detectors.
Good agreement is observed in the central (barrel) region of the detector,
where the differences are at the level of 10\% to 30\%.
In the endcap regions, the simulation
is able to reproduce the shape of the measurements, with a quantitative 
agreement within a factor of two. A similar agreement, within a factor two, 
is observed by the 
CSCs.
The RPC average sensitivities determined with GEANT4 are: neutrons $(0.26\pm 0.03)\%$; 
photons: $(1.6 \pm 0.2)\%$; {\rm e+/e-}: $(35 \pm 16)\%$.

\section{Irradiation tests at the GIF++ facility}
\label{sec:gif}

\begin{figure}[htbp]
\centering 
\includegraphics[width=.6\textwidth]{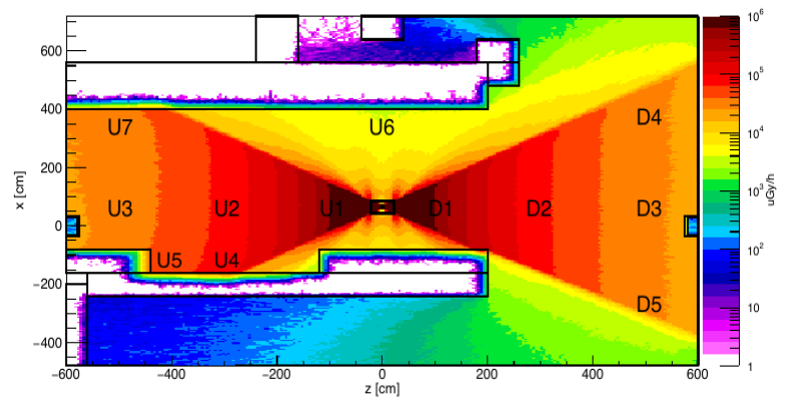}
\caption{\label{fig:gifpp} Dose distribution inside the GIF++
bunker. The labels U (upstream) and D (downstream) refer to
the two independent irradiation regions of the GIF++ bunker,
classified as ``upstream'' or ``downstream'' relative to the 
incoming charged particle beam. The chambers being irradiated 
are located near the positions labelled as D4 (DTs), U3 (RPCs),
D2 (CSCs), and D1 (GEMs).
}
\end{figure}

The expected higher background conditions at the HL-LHC are 
realistically simulated at 
the CERN Gamma Irradiation Facility (GIF++)~\cite{gif}, where beams of
high energy charged particles, mainly 100 GeV muons, are provided 
with photons from an intense (14 TBq) Cs$^{137}$ source. 
The dose distribution inside the GIF++ bunker is shown in 
figure~\ref{fig:gifpp}~\cite{gif}. The energy
of the emitted photons, 662 keV, is within the typical range (0.1 to 10 MeV)
of neutron-induced photons emitted at the LHC.
Accelerated irradiation tests are performed, allowing to study 
the performance and stability, and to asses the 
longevity of detectors expected to cope for several decades
with the HL-LHC conditions, i.e. with doses and rates of the order
of five times the LHC ones (as mentioned in Section~\ref{sec:intro}). 
Effects of prolonged exposition to radiation
may result in loss of gas gain; rise in spurious signal rates (dark rate); 
increase in leakage currents (dark current); development of self-sustained 
discharges set off at high radiation rates (Malter effect). Existing and
new detectors are therefore being certified for the doses and rates
expected at the HL-LHC. Maximum hit rates and charges expected to be accumulated
in the chamber gas volume, relevant for the detector longevity,
are summarized in table~\ref{tab:table1}~\cite{tdr}.

\begin{table}[htbp]
\centering
\caption{\label{tab:table1}
Expected hit rate and accumulated charge for the CMS muon detectors after the Phase-2 upgrades, 
at the end of HL-LHC running, assuming an instantaneous luminosity of  
$5 \cdot 10^{34}$~cm$^{-2}$~s$^{-1}$.
Only the worst-case values in the most exposed chambers of each subdetector are given.   
The numbers shown do not include safety factors.
The column RPC refers to the RPC chambers already present during Phase-1,
while the new chambers to be installed are referred to as iRPCs.
}
\smallskip
\begin{tabular}{|l|ccccccc|}
\hline
 & DT & CSC & RPC & iRPC & GE1/1 & GE2/1 & ME0 \\
$|\eta|$ range & 0-1.2& 0.9-2.4 & 0-1.9 & 1.8-2.4& 1.6-2.15& 1.6-2.4& 2.0-2.8 \\
\hline
Hit rate (Hz/cm$^2$)  & 50& 4500& 200& 700& 1500& 700& 48000 \\
Charge per wire (mC/cm) & 20 & 110 & & & & & \\
Charge per area (mC/cm$^2$) & & &  280& 330& 6& 3& 280\\
\hline
\end{tabular}
\end{table}

\begin{figure}[htbp]
\centering 
\includegraphics[width=.5\textwidth]{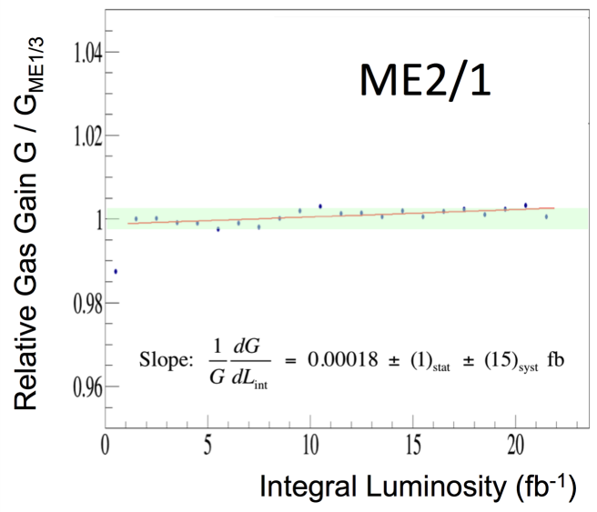}
\caption{\label{fig:gif_csc} 
Relative gas gain in the CSC ME2/1 chambers, the most exposed to radiation background,
normalized to the least exposed ME1/3 chambers, taken as a reference in the comparison.
}
\end{figure}

\begin{figure}[htbp]
\centering 
\includegraphics[width=.5\textwidth]{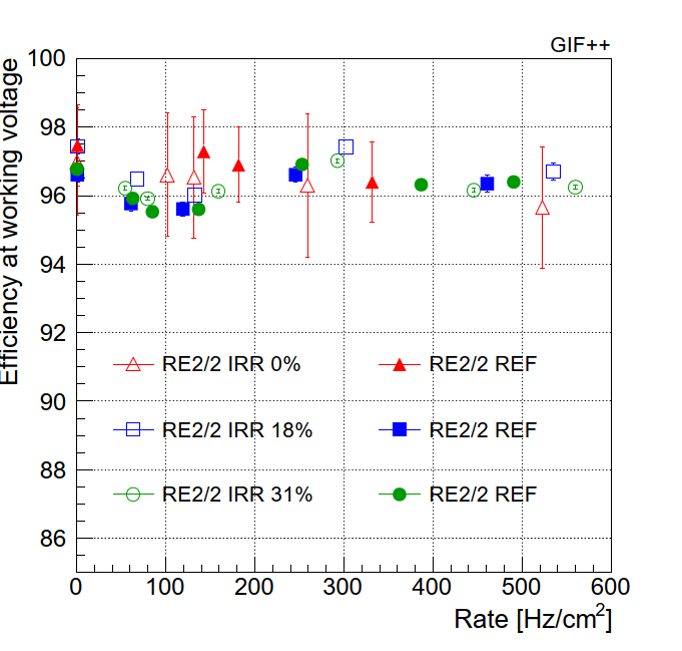}
\caption{\label{fig:fig_rpc} RPC hit efficiency at the working point
as a function of the $\gamma$ background rate per unit area,
for values of the integrated charge up to ~600 mC/cm$^2$, for
both irradiated (IRR)
and non-irradiated reference (REF) chambers.
The measurements are shown at different values (0, 18 and 31\%) of the 
total integrated charge expected at the HL-LHC. }
\end{figure}

\begin{figure}[htbp]
\centering 
\includegraphics[width=.5\textwidth]{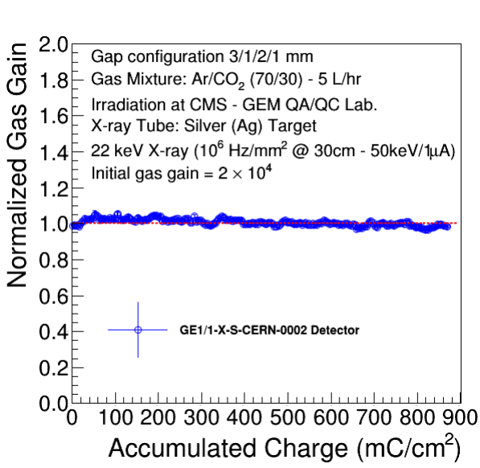}
\caption{\label{fig:fig_gem} 
Normalized gas gain in the GEMs as a function of the accumulated charge.}
\end{figure}

Chambers of all the muon subdetectors are being irradiated
at the GIF++.
We refer to~\cite{isidro} for a detailed discussion of the longevity 
studies performed by the DTs.
The CSCs have collected a total integrated charge of about 330 mC/cm,
corresponding to three times the integrated charge per wire expected
at the HL-LHC (figure~\ref{fig:gif_csc}) with no evidence of gas gain loss.
The average gas gain of the most exposed ME2/1 chambers is normalized to 
the one of the least exposed ME1/3 chambers, used as a reference.
The RPCs (figure~\ref{fig:fig_rpc}) see no noticeable effects of detector 
degradation for values the integrated charge up to ~600 mC/cm$^2$ 
corresponding to roughly 70\% of the ones expected at the HL-LHC.  
Longevity tests are also performed on large size prototype of Phase-2
RPC detectors, showing stability of the main detector parameters.
The GEMs have observed no aging effect with total accumulated charge 
of 125 mC/cm$^2$, corresponding to 10 years of GE1/1 (GE2/1)
operation at the HL-LHC, and about 45\% of the total ME0 operation.
Irradiation tests for ME0 have been performed also with an X-ray source,
allowing to collect an integrated charge of 875 mC/cm$^2$, corresponding
to 10 years of operation (figure~\ref{fig:fig_gem}).
Safety factors of at least 3 are included for all subdetectors.

\section{Conclusions}

Extensive measurements have been performed during the LHC Run1 and Run2 data taking periods,
leading to an accurate understanding of the radiation background in the CMS Muon system.  
The simulation, with FLUKA being mainly used for flux and dose estimation, and GEANT4 for 
sensitivity studies, well describes the data in the central
part of the detector, with differences in the range of 10\% to 30\%. 
The agreement between data and simulation is within a factor 
two in the endcap.
The level of agreement with present measurements, in the regions covered by the existing detectors, 
allows to estimate the expected radiation background where new detectors are going to be 
installed at the HL-LHC, and it is essential for detector design ad future operation.
Longevity tests are being performed at the CERN GIF++ facility. They show no evidence of 
aging effects at the total integrated charge collected so far.

%



\begin{thebibliography}{99}

\bibitem{tdr}
CMS Collaboration, \emph{The Phase-2 Upgrade of the CMS Muon Detectors}, 
CMS-TDR-016 (2017).

\bibitem{cms_det}
CMS Collaboration, \emph{The CMS experiment at the CERN LHC}, JINST {\bf 3} (2008)
S08004.

\bibitem{gem}
F. Sauli, \emph{GEM: A new concept for electron amplification in gas detectors},
Nucl.Instr. Meth. A {\bf 386} (1997) 531.

\bibitem{muon}
CMS Collaboration, \emph{Performance of the CMS muon detector and muon
reconstruction with proton-proton collisions at $\sqrt{s} =$ 13 TeV}, JINST {\bf 13}
(2008) P06015.

\bibitem{fluka1}
A. Ferrari et al., \emph{FLUKA: a multi-particle transport code}, CERN-2005-10
(2005).

\bibitem{fluka2}
G. Battistoni et al., \emph{Overview of the FLUKA code}, Ann. Nucl. Energy {\bf 82}
(2015), 10.

\bibitem{geant1}
GEANT4 Collaboration, \emph{Geant4 - a simulation toolkit}, Nucl.Instr. Meth. A {\bf 506}
(2003) 250.

\bibitem{geant2}
Geant4 Collaboration, \emph{Geant4 developments and applications}, 
IEEE Trans. Nucl. Scien. {\bf 53} (2006) 270.

\bibitem{geant3}
Geant4 Collaboration, \emph{Recent developments in Geant4}, Nucl.Instr. Meth. A {\bf 835}
(2016) 186.

\bibitem{gif}
D. Pfeiffer et al., \emph{The radiation field in the Gamma Irradiation Facility (GIF++)
at CERN}, Nucl.Instr. Meth. A {\bf 866} (2017) 91.

\bibitem{isidro}
G. Abbiendi et al.,, \emph{Study of the effects of radiation on the CMS Drift 
Tubes Muon Detector for the HL-LHC}, 
JINST {\bf 14} (2019) C12010.

%
%





\end{thebibliography}
\end{document}